\documentclass[letterpaper]{article}
\usepackage{amsmath}

\input{tcilatex}
\begin{document}

\title{\hfill {\normalsize UMTG - 14}\bigskip \\
Associativity, Jacobi, Bremner, and All That}
\author{Thomas Curtright \\
Department of Physics, University of Miami, Coral Gables, FL 33124\bigskip \\
\textit{Talk presented at QTS6, University of Kentucky, 21 July 2009,}\\
\textit{and at Miami 2009, Fort Lauderdale, Florida, 20 December 2009.}\\
\textit{To appear in Journal of Physics: Conference Series.}}
\date{}
\maketitle

\begin{abstract}
I discuss various aspects of multi-linear algebras related to associativity.
\end{abstract}

\section*{Introduction}

Nambu introduced a multilinear operator $N$-bracket in the context of a
novel formulation of mechanics \cite{Nambu}:%
\begin{equation}
\left[ A_{1}A_{2}\cdots A_{N}\right] =\sum_{\sigma \in S_{N}}\limfunc{sgn}%
\left( \sigma \right) ~A_{\sigma _{1}}\cdots A_{\sigma _{N}}\ ,
\end{equation}%
where the sum is over all $N!$ permutations of the operators. \ For example,
the operator 3-bracket is 
\begin{equation}
\left[ ABC\right] =ABC-ACB+BCA-BAC+CAB-CBA\ .
\end{equation}%
The operator product here is assumed to be \emph{associative}.

The same construction independently appeared in the mathematical literature
more than 50 years ago \cite{Higgins,Kurosh}. \ The theory of such
multi-operator products, as well as their \textquotedblleft classical
limits\textquotedblright\ in terms of multivariable Jacobians, has been
studied extensively \cite%
{Bremner,CFJMZ,CFZ,Curtright,deAzcarraga,Devchand,Dito,Filippov,Gautheron,Hanlon,Lada,Pojidaev,Schlesinger,Takhtajan,Vainerman,Vaisman}%
.

From an algebraic point of view, it is natural to seek \emph{the analogue of
the Jacobi identity} for operator $N$-brackets. \ For the case of even $N$%
-brackets, the obvious generalization where one $N$-bracket acts on another
leads to a true identity. \ However, for odd $N$-brackets this usually does 
\emph{not} work. \ For instance, it is almost always true that%
\begin{equation}
\left[ \left[ ABC\right] DE\right] -\left[ \left[ ADE\right] BC\right] -%
\left[ A\left[ BDE\right] C\right] -\left[ AB\left[ CDE\right] \right] \neq
0\ .
\end{equation}%
That is to say, the so-called FI (\textquotedblleft fundamental
identity\textquotedblright ) \emph{fails}. \ There is one especially notable
exception: \ $su\left( 2\right) $, as described by Nambu \cite{Nambu}.

Fortunately, even brackets are not odd. \ They need only act twice to yield
an identity. \ Namely \cite{deAzcarraga,Hanlon}, 
\begin{equation}
\left[ ~B_{1}\cdots B_{N-1}~\left[ B_{N}\cdots B_{2N-1}\right] ~\right] =0%
\text{ \ \ for }N\text{ even.}
\end{equation}%
Here, total antisymmetrization of all the $B$s is understood. \ When $N=2$
this is the familiar Jacobi identity. \ The proof is by direct calculation
and follows as a consequence of associativity.

Unfortunately, an odd $N$-bracket acting on just one other odd $N$-bracket
does \emph{not} vanish even when totally antisymmetrized over all entries,
but rather produces a $\left( 2N-1\right) $-bracket \cite%
{deAzcarraga,Curtright}. \ Therefore the simplest identity obeyed by odd
brackets of only one type, that does not introduce higher-order brackets,
requires that they act at least \emph{thrice}.

\section*{Bremner Identity and GBIs}

Bremner \cite{Bremner} proved an identity (henceforth the \textquotedblleft
BI\textquotedblright ) for associative operator 3-brackets acting thrice.%
\begin{equation}
\left[ ~\left[ A~\left[ bcd\right] ~e\right] ~f~g\right] =\left[ ~\left[ Abc%
\right] ~\left[ def\right] ~g\right] \ ,
\end{equation}%
where it is understood that all \emph{lower} case entries are totally
antisymmetrized by implicitly summing over all $6!=720$ \emph{signed}
permutations of them.

The BI can be proven through a resolution of both LHS and RHS as a series of
canonically ordered words. \ By direct calculation we find%
\begin{equation}
\left[ \left[ A\left[ bcd\right] e\right] fg\right]
=24~Abcdefg-36~bAcdefg+36~bcAdefg-24~bcdAefg+36~bcdeAfg-36~bcdefAg+24~bcdefgA\ .
\label{3on3on3Resolution}
\end{equation}%
The same expansion holds for $\left[ \left[ Abc\right] \left[ def\right] g%
\right] $, again by direct calculation. \ That is to say, both $\left[ \left[
A\left[ bcd\right] e\right] fg\right] $ and $\left[ \left[ Abc\right] \left[
def\right] g\right] $ can be rendered as a 7-bracket plus another 3-bracket
containing 3-brackets:%
\begin{equation}
\left[ \left[ A\left[ bcd\right] e\right] fg\right] =\frac{1}{20}~\left[
Abcdefg\right] -\frac{1}{6}~\left[ A\left[ bcd\right] \left[ efg\right] %
\right] =\left[ \left[ Abc\right] \left[ def\right] g\right] \ .
\end{equation}%
Thus the BI amounts to the combinatorial statement that there are two
distinct ways to write a 7-bracket in terms of nested 3-brackets.

Xiang Jin, Luca Mezincescu, and I proved that a similar identity holds for
any odd-order bracket acting thrice \cite{CJM}. \ For odd $N=2L+1$, this
generalized BI (\textquotedblleft GBI\textquotedblright ) is 
\begin{equation}
\left[ ~\left[ AB_{1}\cdots B_{2L}\right] ~\left[ B_{2L+1}\cdots B_{4L+1}%
\right] ~B_{4L+2}\cdots B_{6L}~\right] =\left[ ~\left[ A~\left[ B_{1}\cdots
B_{2L+1}\right] ~B_{2L+2}\cdots B_{4L}\right] ~B_{4L+1}\cdots B_{6L}~\right]
\ .
\end{equation}%
Again, this identity is a consequence of only associativity. \ Thus all odd
brackets built from associative products of operators need only act thrice
to yield an identity. \ Given an hypothesized closed algebra of odd $N$%
-brackets, the GBI provides the simplest test for consistency with an
underlying associative product.

To prove the GBI, we again expanded in terms of canonically ordered words. \
By direct calculation,%
\begin{equation}
\left[ \left[ A\left[ B_{1}\cdots B_{2L+1}\right] ~B_{2L+2}\cdots B_{4L}%
\right] ~B_{4L+1}\cdots B_{6L}~\right] =\sum_{n=0}^{6L}\left( -1\right)
^{n}m_{n}~B_{1}\cdots B_{n}~A~B_{n+1}\cdots B_{6L}\ ,
\end{equation}%
where it is implicit that one is to totally antisymmetrize over all the $B$%
s. \ All the coefficients in the resolution are integers. \ Explicitly,%
\begin{eqnarray}
m_{n} &=&\left( 2L+1\right) !\left( 2L\right) !\left( 2L-1\right) !\times
c_{n}\ ,  \notag \\
c_{n} &=&\left\{ 
\begin{array}{ll}
\left( n+1\right) \left( 4L-n\right) /2 & \text{for \ \ }0\leq n\leq 2L \\ 
10L^{2}-6Ln+L+n^{2} & \text{for \ \ }2L+1\leq n\leq 3L \\ 
c_{6L-n} & \text{for \ \ }3L+1\leq n\leq 6L%
\end{array}%
\right. \ .
\end{eqnarray}%
The same expansion holds for $\left[ \left[ AB_{1}\cdots B_{2L}\right] ~%
\left[ B_{2L+1}\cdots B_{4L+1}\right] ~B_{4L+2}\cdots B_{6L}\right] $, again
by direct calculation. \ Hence the GBI.

For example, $L=1$ gives the previous coefficients (\ref{3on3on3Resolution}%
), while $L=2$ gives 
\begin{equation}
\left( 
\begin{array}{c}
m_{0} \\ 
m_{1} \\ 
m_{2} \\ 
m_{3} \\ 
m_{4} \\ 
m_{5} \\ 
m_{6} \\ 
m_{7} \\ 
m_{8} \\ 
m_{9} \\ 
m_{10} \\ 
m_{11} \\ 
m_{12}%
\end{array}%
\right) =5!4!3!\times \left( 
\begin{array}{c}
c_{0} \\ 
c_{1} \\ 
c_{2} \\ 
c_{3} \\ 
c_{4} \\ 
c_{5} \\ 
c_{6} \\ 
c_{7} \\ 
c_{8} \\ 
c_{9} \\ 
c_{10} \\ 
c_{11} \\ 
c_{12}%
\end{array}%
\right) =5!4!3!\times \left( 
\begin{array}{c}
4 \\ 
7 \\ 
9 \\ 
10 \\ 
10 \\ 
7 \\ 
6 \\ 
7 \\ 
10 \\ 
10 \\ 
9 \\ 
7 \\ 
4%
\end{array}%
\right) =\left( 
\begin{array}{c}
69\,120 \\ 
120\,960 \\ 
155\,520 \\ 
172\,800 \\ 
172\,800 \\ 
120\,960 \\ 
103\,680 \\ 
120\,960 \\ 
172\,800 \\ 
172\,800 \\ 
155\,520 \\ 
120\,960 \\ 
69\,120%
\end{array}%
\right) \ .
\end{equation}%
For simplicity, I emphasize the 3-bracket case in the following.

\section*{3-Algebras}

For a 3-algebra with linearly independent operators $T_{a}$ that obey%
\begin{equation}
\left[ T_{a}T_{b}T_{c}\right] =i~F_{abc}^{\ \ \ d}~T_{d}\ ,
\end{equation}%
the BI becomes, with implicit total antisymmetrization of the six $b_{j}$
indices, 
\begin{equation}
F_{ab_{1}b_{2}}^{\ \ \ x}F_{b_{3}b_{4}b_{5}}^{\ \ \ y}F_{xyb_{6}}^{\ \ \
z}=F_{b_{1}b_{2}b_{3}}^{\ \ \ x}F_{axb_{4}}^{\ \ \ y}F_{yb_{5}b_{6}}^{\ \ \
z}\ .
\end{equation}%
Alternatively, after renaming and cycling indices,%
\begin{equation}
F_{b_{1}b_{2}b_{3}}^{\ \ \ x}\left( F_{axb_{4}}^{\ \ \ y}F_{yb_{5}b_{6}}^{\
\ \ z}-F_{ab_{4}b_{5}}^{\ \ \ y}F_{yxb_{6}}^{\ \ \ z}\right) =0\ .
\label{FTrilinearCondition}
\end{equation}%
This \emph{trilinear relation} is a condition on the structure constants
required by an underlying associativity for any posited 3-algebra.\medskip

\noindent \textbf{Exercise }(25 points; show all details;\ due Friday)%
\textbf{:} \ Use (\ref{FTrilinearCondition}) to prove a classification
theorem for 3-algebras.\medskip

To be more specific, consider now any \emph{closed bilinear algebra} where
all commutators and anticommutators are also elements of the algebra, as
given by%
\begin{equation}
\left[ T_{a}T_{b}\right] =if_{ab}^{\ \ c}~T_{c}\ ,\ \ \ \left\{
T_{a}T_{b}\right\} =g_{ab}^{\ \ c}~T_{c}\ .
\end{equation}%
For example, for $u\left( N\right) $ with the $T_{a}$ given by $N\times N$
matrices, the second RHS involves the well-known $d_{ab}^{\ \ c}$ symbol, as
well as Kronecker delta terms. \ Or, with a bit of freedom of
interpretation, one may think of the operator product expansion of any CFT
in this way.

For a bilinear algebra of this form, the corresponding 3-algebra is also
completely determined, or \textquotedblleft induced.\textquotedblright\ \
This follows from%
\begin{equation}
2\times \left[ ABC\right] =\left\{ \left[ AB\right] C\right\} +\left\{ \left[
BC\right] A\right\} +\left\{ \left[ CA\right] B\right\} \ .
\end{equation}%
The induced 3-algebra structure constants are given in terms of the $f$ and $%
g$ symbols by%
\begin{equation}
2~F_{abc}^{\ \ \ x}=f_{ab}^{\ \ u}~g_{uc}^{\ \ x}+f_{bc}^{\ \ u}~g_{ua}^{\ \
x}+f_{ca}^{\ \ u}~g_{ub}^{\ \ x}\ .
\end{equation}%
Thus the BI conditions on the induced 3-algebra structure constants can be
re-expressed in terms of $f$ and $g$. \ Again with implicit
antisymmetrizations, the BI conditions become%
\begin{equation}
f_{b_{1}b_{2}}^{\ \ u}~g_{ub_{3}}^{\ \ x}\left( F_{axb_{4}}^{\ \ \
y}F_{yb_{5}b_{6}}^{\ \ \ z}-F_{ab_{4}b_{5}}^{\ \ \ y}F_{yxb_{6}}^{\ \ \
z}\right) =0\ .
\end{equation}%
These conditions are indeed obeyed when the $f$ and $g$ symbols satisfy the
conditions wrought by associativity.

The Jacobi identity (JI),%
\begin{equation}
\left[ A\left[ BC\right] \right] +\left[ B\left[ CA\right] \right] +\left[ C%
\left[ AB\right] \right] =0\ ,
\end{equation}%
is a consequence of associativity but it is \emph{not} equivalent to it,
even when augmented with the super Jacobi identity (SJI),%
\begin{equation}
\left[ \left\{ AB\right\} C\right] =\left\{ A\left[ BC\right] \right\}
+\left\{ B\left[ AC\right] \right\} \ .
\end{equation}%
Here, we have used the usual (anti)commutator notation, sans commas, $%
\left\{ AB\right\} =AB+BA$ and $\left[ AB\right] =AB-BA$.

However, there is \emph{another} trilinear identity which, when paired with
the SJI, \emph{is} equivalent to associativity. \ Namely, 
\begin{equation}
\left[ A\left[ BC\right] \right] =\left\{ \left\{ AB\right\} C\right\}
-\left\{ \left\{ AC\right\} B\right\} \ .
\end{equation}%
For want of a more compelling name, we will refer to this third relation as
the \textquotedblleft super-duper Jacobi identity" (SDJI)\footnote{%
Apologies to Irving Berlin.}. \ Note that the JI follows from the SDJI.

For a closed bilinear algebra the SJI and SDJI identities require the
following conditions to be obeyed by the structure constants:%
\begin{equation}
g_{ab}^{\ \ u}~f_{uc}^{\ \ x}=g_{au}^{\ \ x}~f_{bc}^{\ \ u}+g_{bu}^{\ \
x}~f_{ac}^{\ \ u}\ ,\ \ \ f_{bc}^{\ \ u}~f_{au}^{\ \ x}=g_{ab}^{\ \
u}~g_{uc}^{\ \ x}-g_{ac}^{\ \ u}~g_{ub}^{\ \ x}\ .
\end{equation}%
The more familiar JI conditions on the bilinear algebra structure constants
follow from the second of these.

As I mentioned already, the BI conditions for the induced 3-algebra
structure constants, expressed in terms of $f$ and $g$, follow from these
two conditions. \ Moreover, the structure constants $F$ for \emph{any
induced N-bracket algebra} can be expressed in terms of $f$ and $g$ for such
closed bilinear algebras, and it can be shown that the conditions on $F$
imposed by associativity are indeed satisfied as a consequence of these same
two conditions on $f$ and $g$.

As an \emph{aside}, there are deformed versions of these identities
involving the \textquotedblleft quommutators\textquotedblright\ 
\begin{equation}
\left[ AB\right] _{\lambda }\equiv \lambda AB-\lambda ^{-1}BA\ .
\end{equation}%
These naturally lead to 3-brackets. \ For example, the \textquotedblleft
Jaquobi identity\textquotedblright :%
\begin{equation}
\left[ \left[ AB\right] _{\lambda }C\right] _{\mu }+\left[ \left[ BC\right]
_{\lambda }A\right] _{\mu }+\left[ \left[ CA\right] _{\lambda }B\right]
_{\mu }=\tfrac{1}{2}\left( \lambda +\lambda ^{-1}\right) \left( \mu -\mu
^{-1}\right) \left[ ABC\right] +\tfrac{1}{2}\left( \lambda -\lambda
^{-1}\right) \left( \mu -\mu ^{-1}\right) \left\{ ABC\right\} \ .
\end{equation}%
Etc.\footnote{%
The \emph{ultimate} \emph{Jaquobi identity} would involve six complex
parameters, $\lambda _{j}$ and $\mu _{j}$, $j=1,2,3$, as in: \ $\left[ \left[
AB\right] _{\lambda _{1}}C\right] _{\mu _{1}}+\left[ \left[ BC\right]
_{\lambda _{2}}A\right] _{\mu _{2}}+\left[ \left[ CA\right] _{\lambda _{3}}B%
\right] _{\mu _{3}}$. \ Requiring that this vanish gives six equations for
the parameters. \ Assuming $\mu _{1}\neq 0,\ \lambda _{3}\neq 0,\ \mu
_{3}\neq 0$, the \emph{generic} solution is: $\lambda _{1}=\mu _{1}\lambda
_{3}\mu _{3}$, $\ \lambda _{2}=\mu _{1}\lambda _{3}$, and $\mu _{2}=\frac{1}{%
\mu _{1}\mu _{3}}$. \ So the solution manifold has complex dimension three,
including an overall complex scale. \ For fixed scale, it is in fact a
geometrically \emph{ruled surface}, and it must contain the usual Jacobi,
the super Jacobi, and the super-duper Jacobi identities. \ Indeed, it does.
\ The symmetric group is a symmetry of the solution manifold, so other
solutions are given by permutations of $1,2,3$. \ There is also parity: \
Another solution is obtained from the generic one just by flipping the signs
of all the parameters.} \ Here, we have also used the totally symmetrized
3-bracket: \ $\left\{ ABC\right\} =ABC+ACB+BCA+BAC+CAB+CBA$.

\section*{Classical Manifolds}

There are also interesting questions for classical manifold theory that
arise in this context. \ A \emph{classical} 3-bracket is defined by%
\begin{equation}
\left[ A,B,C\right] =\omega ^{abc}~\partial _{a}A~\partial _{b}B~\partial
_{c}C\ ,
\end{equation}%
with antisymmetric but otherwise arbitrary 3-tensor $\omega ^{abc}$. \ The
combination that constitutes the so-called FI is%
\begin{gather}
\left[ E,F,\left[ A,B,C\right] \right] -\left[ \left[ E,F,A\right] ,B,C%
\right] -\left[ A,\left[ E,F,B\right] ,C\right] -\left[ A,B,\left[ E,F,C%
\right] \right]  \notag \\
=\left( \omega ^{abc}\omega ^{def}-\omega ^{dbc}\omega ^{aef}-\omega
^{adc}\omega ^{bef}-\omega ^{abd}\omega ^{cef}\right) ~\partial _{d}\left(
\partial _{a}A~\partial _{b}B~\partial _{c}C~\partial _{e}E~\partial
_{f}F\right)  \notag \\
+\left( \partial _{a}A~\partial _{b}B~\partial _{c}C~\partial _{e}E~\partial
_{f}F\right) \left( \omega ^{def}\partial _{d}\omega ^{abc}-\omega
^{dbc}\partial _{d}\omega ^{aef}-\omega ^{adc}\partial _{d}\omega
^{bef}-\omega ^{abd}\partial _{d}\omega ^{cef}\right) \ .
\end{gather}%
In the literature, when $\omega $ is such that this vanishes, this is called
a \emph{Nambu-Poisson manifold}. \ This gives two types of bilinear
constraints on $\omega $, obviously.

But, to conclude this talk, it seems more reasonable (to me at least) that
one should impose, instead of the FI, a classical analogue of the BI. \ For
a classical N-bracket involving $n\geq N$ (odd) independent variables, with
antisymmetric but otherwise arbitrary N-tensor $\omega ^{a_{1}\cdots a_{N}}$,%
\begin{equation}
\left[ B_{i_{1}},B_{i_{2}},\cdots ,B_{i_{N}}\right] =\omega ^{a_{1}\cdots
a_{N}}~\partial _{a_{1}}B_{i_{1}}~\cdots ~\partial _{a_{N}}B_{i_{N}}\ ,
\end{equation}%
we define a \emph{Bremner-Poisson manifold} as one for which the BI holds. \
This leads to requirements on the $\omega $ tensor that differ from those
imposed by the FI. \ So defined, Bremner-Poisson and Nambu-Poisson manifolds
are different, in general. \ We will discuss this in more detail elsewhere.

Perhaps N-brackets and algebras have an important role to play in physics,
as originally suggested by Nambu. \ Recently there has been considerable
interest in $N$-brackets, especially $3$-brackets, as expressed in the
physics literature (see \cite{CFJMZ}\ and references therein). \ These ideas
await further development.

\textbf{Acknowledgments} \ \textit{I thank the organizers of this conference
for the opportunity to talk about this work, and I thank my collaborators,
David Fairlie, Xiang Jin, Luca Mezincescu, and Cosmas Zachos, for sharing
their thoughts about Nambu brackets. \ This work was supported by NSF Awards
0555603 and 0855386.}

\end{document}